# Aryabhata's Mathematics

Subhash Kak

## Introduction

This paper presents certain aspects of the mathematics of Āryabhaṭa (born 476) that are of historical interest to the computer scientist. Although Āryabhaṭa is well known in history of astronomy, the significance of his work to computing science is not as well appreciated and one of the objectives of the 2006 RSA Conference was to address this lacuna.

The cryptographic community has recently become aware of Āryabhaṭa because his algorithm to solve congruences can be used in place of the Chinese Remainder Theorem in certain situations [33]. His code has had applications in securing passwords and personal identification numbers [37],[38].

Very little is known of the personal life of Āryabhaṭa, and the summary that follows is based on my essays in Scribner's *Encyclopedia of India* [1]. He was born in Aśmaka but later lived in Kusumapura, which the commentator Bhāskara I (c. 629) identifies with Pāṭaliputra (modern Patna). It appears that he was the *kulapati* (head) of the University at Nālandā in Magadha [2].

There is no agreement about the location of Aśmaka, but since the Āryabhaṭian school has remained strong in Kerala in South India, some believe that he may have belonged to that region. In addition, Āryabhaṭa employed the Kali era (3102 BC) which remains popular in South India. The Aśmaka people were taken in early literature to belong either to northwest India or the region in South India between the rivers Godāvarī and Narmadā.

Āryabhaṭa wrote at least two books: the *Āryabhaṭīya* and the *Āryabhaṭa-siddhānta*, of which the latter is known only through references in other works. The *Āryabhaṭīya* is one of the landmarks of the history of astronomy. Apart from its continuing influence on Indian astronomy, it was translated into Arabic as *Arajbahara,* which in turn influenced Western astronomers. The commentaries on his work by Bhāskara I, Someśvara (11[th] or 12[th] century), Parameśvara, Nīlakaṇṭha Somayāji (15[th] century) and others provide invaluable help in understanding Āryabhaṭa's mathematics [2],[3]. They also provide us important facts about the symbolic notations that were used in Indian mathematics.

Āryabhaṭa's main contributions to mathematics include the good approximation of 3.1416 for $\pi$, a table of sine-differences, and root-extraction methods. He also presented a method to solve indeterminate equations of a certain type that are important in astronomy and computer science, and he used a novel word code to represent numbers.

From the point of view of physics, his ideas on relativity of motion and space [30] prefigure that of Galileo. He took the earth to be spinning on its axis and he gave the periods of the planets with respect to the sun.

His figure for the sidereal rotation of the earth was extremely accurate. He made important innovations in planetary computations by using simplifying hypotheses, and he presented a method of finding the celestial latitudes of the planets.

Āryabhaṭa's work is best understood by examining the Indian scientific tradition to which he belonged: for general reviews of this tradition, see [4]-[8]; and for a broader introduction to the earliest Indian science, see [9]-[13]. For an even broader context, the beginnings of Indian writing and number systems are summarized in [14]-[20],[32].

This paper will first review some recent theories about the originality of Āryabhaṭa's ideas. Then it will summarize his main work, the *Āryabhaṭīya*, present an overview of his relativity, his algorithm to solve linear indeterminate equations, and finally his word code for numbers. It is meant to introduce the reader to the topics and for deeper analysis other references should be examined. For Āryabhaṭa's root extraction methods, see [21],[22], in which it is shown that these methods are different from those of the Greeks.

## Āryabhaṭa and his Predecessors

At the end of his own text, Āryabhaṭa thanks Brahmā (Brahma or Paitāmaha Siddhānta), indicating that he was connected to an existing tradition. An early version of the Paitāmaha Siddhānta is preserved in the sixth century summary of five siddhāntas by Varāhamihira.

Some scholars have suggested that Āryabhaṭa and other Indian astronomers borrowed certain mathematical techniques and observations from Greek and Babylonian astronomy [23],[24]. The text that is used for comparison is the Almagest, the twelfth century Arabic version of Ptolemy's astronomical text of which the original Greek text is lost. This late Arabic text which was later translated back into Greek is bound to have an accretion of Islamic material, which is especially true of the sections concerning star locations that were given much attention by Islamic astronomers. As a point of comparison, the Sūrya Siddhānta of which we have a summary from sixth century by Varāhamihira is quite different from the later version that has come down to us.

In the revisionist view of Indian astronomy, elements of the Indian texts of the first millennium AD are taken to be borrowed from a text that dates only from 12[th] century AD. Critics have argued that this is an example of the Eurocentric view that asserts science arose only in Greece and Europe, with the Babylonians credited with accurate observations, and any novel scientific models encountered outside of this region are taken to be borrowed from the Greeks. If evidence in the larger Greek world for a specific scientific activity is lacking then the presence of it outside that region is termed a remnant of Greco-Babylonian science. Such material is gathered together in what is called "recovery of Greco-Babylonian science."

As a specific example, consider the Rsine table given by Āryabhaṭa. It was proposed by Toomer [23] that this table was derived from the Table of Chords of Hipparchus. But as explained by Narahari Achar [26], there is no proof that such a table ever existed. G.J. Toomer, a modern translator of the Greek translation of the Arabic version of the Almagest, nevertheless proceeded to invent such a table to explain two ratios relating to the moon's orbit.

Toomer's construction of the non-existent Table of Chords of Hipparchus is ultimately unsuccessful. His construction does not lead to the numbers actually used by Ptolemy. Here's how Achar summarizes the process [26]: "Since the numbers so obtained do not agree with those given by Ptolemy (and ascribed to Hipparchus), the natural conclusion should have been that there is no relationship between the numbers of Hipparchus and those derived from Āryabhaṭa. However, in his zeal to prove the non-originality and the indebtedness of Āryabhaṭa to Hipparchus, Toomer further hypothesizes a particular mistake to have been committed by Hipparchus. Even after all this, still there is no agreement between the two sets of numbers. Yet, Toomer offers this as the conclusive proof that Āryabhaṭa borrowed from Hipparchus..!"

A similar case has also been made for Āryabhaṭa's planetary constants and once again the numbers do not match up [25]. Achar also points out that Hayashi [27] has demonstrated the originality of Āryabhaṭa's Rsine table. The lesson to be drawn from the above is that it is foolhardy to build elaborate theories of transmission of scientific ideas in the absence of evidence.

## The *Āryabhaṭīya*

The *Āryabhatīya* is divided into four parts. The first part, *Gītikā*, provides basic definitions and important astronomical parameters. It mentions the number of rotations of the earth and the revolutions of the sun, moon, and the planets in a period of 4,320,000 years. This is a partially heliocentric system because it presents the rotation information of the planets with respect to the sun. The second part, *Gaṇita*, deals with mathematics. The third part, *Kālakriyā,* deals with the determination of the true position of the sun, the moon, and the planets by means of eccentric circles and epicycles. The fourth part, *Gola*, deals with planetary motions on the celestial sphere and gives rules relating to various problems of spherical astronomy.

The notable features of the *Āryabhatīya* are Āryabhaṭa's theory of the earth's rotation, and his excellent planetary parameters based on his own observations made around 512 A.D. which are superior to those of others [3]. He made fundamental improvements over the prevailing *Sūrya-siddhānta* techniques for determining the position of planets. Unlike the epicycles of Greek astronomers, which remain the same in size at all places, Āryabhaṭa 's epicycles vary in size from place to place.

Āryabhaṭa took the sun, the moon, and the planets to be in conjunction in zero longitude at sunrise at Laṅkā on Friday, 18 February, 3102 BC. In a period of 4.32 million years, the moon's apogee and ascending node too are taken to be in conjunction with the planets. This allowed him to solve various problems using whole numbers.

The theory of planetary motion in the *Āryabhaṭīya* is based on the following ideas: the mean planets revolve in geocentric orbits; the true planets move in eccentric circles or in epicycles; planets have equal linear motion in their respective orbits. The epicycle technique of Āryabhaṭa is different from that of Ptolemy and it appears to be derived from an old Indian tradition.

Āryabhaṭa made important innovations on the traditional Sūrya-siddhānta methods for the computation of the planetary positions. The earlier methods used four corrections for the superior planets and five for the inferior planets; Āryabhaṭa reduced the number of corrections for the inferior planets to three and improved the accuracy of the results for the superior planets.

Āryabhaṭa considers the sky to be 4.32 million times the distance to the sun. He and his followers believed that beyond the visible universe illuminated by the sun and limited by the sky is the infinite invisible universe. Rather than taking the universe to be destroyed at the end of the Brahmā's day of 4.32 billion years, he took the earth to go through expansion and contraction equal to one *yojana*.

The measures used in ancient India are summarized in the table below [35]:

| Measure | aṅgulas | centimeters |
| --- | --- | --- |
| *aṅgula* | 1 | 1.763 |
| *vitasti* (*tāla*) | 12 | 21.156 |
| *pāda* | 14 | 24.682 |
| *aratni, P-hasta* | 24 | 42.312 |
| *daṇḍa* | 96 | 169.248 |
| *dhanus* | 108 | 190.404 |

According to the Arthaśāstra, one *yojana* equals 8,000 *dhanus,* whereas Āryabhaṭa took it to be 8,000 *nṛ* (height of man). Āryabhaṭa defines his *nṛ* to be equal to 96 *aṅgula*, so that it is the *daṇḍa* of the table above. Thus his *yojana* was 8/9 of the Arthaśāstra measure. In other words, the Āryabhaṭa *yojana* was 13.54 km. As he gives the diameter of earth to be 1,050 *yojana,* his estimate exceeds the modern value by 11 percent.

Āryabhaṭa was clear about the relativity of motion: "Just as a man in a boat moving forward sees the stationary objects (on the shore) as moving backward, just so are the stationary stars seen by the people on earth as moving exactly towards the west." He described relativity of space thus: "Heavens and the Meru mountain are at the centre of the land (i.e., at the north pole); hell and the Badavāmukha are at the center of the water (i.e., at the south pole). The gods (residing at the Meru mountain) and the demons (residing at the Badavamukha) consider *themselves positively and permanently below each other."* Thus he addressed both the relativity of motion and space much before Galileo [30].

## Kuṭṭaka -- The Pulverizer

This algorithm from the Āryabhaṭiya, called the *kuṭṭaka* or simply the *pulverizer*, is to solve the indeterminate (Diophantine) equation

$$a x + c = b y \qquad (1)$$

We do not know if Āryabhaṭa devised this himself or he was including an already existing method only because it is useful in astronomical calculations. In recent literature, this has also been called the Āryabhaṭa algorithm [28],[33],[34].

As background, note that symbolic algebra has a long history in India. Thus early texts show how the equation *10 y – 8 = y² + 1* was written down in the following manner:

*yāva 0  yā 10  rū $\bar{8}$*
*yāva 1  yā 0   rū 1*

where *yā* stands for the unknown variable *yāvattāvat*, *yāva* means the square of the unknown variable, and *rū* is short for *rūpa*, the absolute term, and $\bar{8}$ means "-8" [3, page lxxvii]. Writing one expression below another meant equality, so the given equation could be written in the modern form as:

*yāva 0  yā 10  rū $\bar{8}$ = yāva 1  yā 0  rū 1*

Bhāskara explains that one could speak of two kinds of *kuṭṭākāra:* one with residue, and the other without residue. Āryabhaṭa also speaks of both these cases; he uses the term *gulikā* in place of *yāvattāvat* for the unknown variable.

*A historical note:* The *kuṭṭaka* may be used in place of the Chinese Remainder Theorem to find the multiplicative inverse in the solution of congruences. According to the historian of mathematics Yoshio Mikami (quoted in [3, page 311]), the Chinese Remainder Theorem was developed in a general form only by I-tsing in 727, after its initial statement in the third-century book by Sun Tzu. I-tsing was a Sanskrit scholar who visited India in 673 and he brought the mathematics of *kuṭṭaka* to China from India for help in calendrical calculations.

For the residual pulverizer, the *kuṭṭaka* algorithm (*Gaṇita*, verses 32-33) is given as below:

> Divide the divisor corresponding to the greater remainder by the divisor corresponding to the smaller remainder. (Discard the quotient).
> Divide the remainder obtained (and the divisor) by one another (until the number of quotients of the mutual divisions in even and the final remainder is small enough).
> Multiply the final remainder by an optional number and to the product obtained add the difference of the remainders (corresponding to the greater and smaller divisors); then divide the sum by the last divisor of the mutual division. The optional number is to be so chosen that this division is exact.
> Now place the quotients of the mutual division one below the other in a column; below them right the optional number and underneath it the quotient just obtained. Then reduce the chain of numbers which have been written down one below the other, as follows:
> Multiply by the last but one number (in the bottom) the number just above it and then add the number just below it (and then discard the lower number).
> Repeat this process until there are only two numbers in the chain.
> Divide (the upper number) by the divisor corresponding to the smaller remainder, then multiply the remainder obtained by the divisor corresponding to the greater remainder, and then add the greater remainder: the result is the *dvicchedāgra* (i.e., the number answering to the two divisors). (This is also the remainder corresponding to the divisor equal to the product of the two divisors).
> [2, page 75]

Consider the special case when $c = 1$ in equation (1), we have the special case

$$a x + 1 = b y \qquad (2)$$

This implies that
$$a^{-1} = -x \bmod b \qquad (3)$$
and
$$b^{-1} = y \bmod a \qquad (4)$$

**Example:** Consider $137 x + 10 = 60 y$

We first find the gcd of 137 and 60

```
60 )137(2
    120
     17 )60( 3
         51
          9)17(1
            9
            8 )9(1
               8
```

1     Also, $1 \times 18 - 10 = 8 \times 1$

We now construct the Āryabhaṭa Array

```
2   2   2    2     297 = y
3   3   3   130    130 = x
1   1   37   37
1   19  19
18  18
1
```

Therefore, for $137x + 10 = 60y$

    $x = 130$ and $y = 297$

    *297* mod *137* = *23* and *130* mod *60* = *10*

    *137* $\times$ *10* + *10* = *60* $\times$ *23* = *1380*

The multiplicative inverse mod m for *1/137* mod *60* and *1/60* mod *137*

    *137 x + 1 = 60 y*

The Āryabhaṭa array may be compressed as (137, 60) =1

```
2   16 = y
3   7 = x
1   2
1   1    137⁻¹ = -7 mod 60 = 53
1             60⁻¹ = 16 mod 137
```
(rendered: $137^{-1} = -7 \bmod 60 = 53$, $60^{-1} = 16 \bmod 137$)

For comparison with the Extended Euclid Algorithm, see Rao and Yang [33].

We can also see if the solution to the indeterminate equations:

    *x* mod *60* = *0*
    *x* mod *137* = *10*

    is *x = 1380*

And the solution of:

    *x* mod *60* = *5*
    *x* mod *137* = *15*

is simply  *x = 1380 + 5 = 1385*

## Substitution Codes

Before considering Āryabhaṭa's word substitution code for numbers, it is good to first look at the Indian tradition of cryptography and representation of numbers by words.

David Kahn, in his *The Codebreakers* [31] has a brief section on Indian cryptography. Vātsyāyana's *Kāma-sūtra* describes cryptographic writing as *mlecchita-viklapa*. In his commentary on the *Kāma-sūtra*, Yaśodhara describes two kinds of this writing:

1. *Kauṭiliyam.* In this letter substitutions are based on phonetic relations, such as vowels becoming consonants. A simplification of this form is called *durbodha*.
2. *Mūladeviya*. Its cipher alphabet consists of pairing letters and using the reciprocal ones:

    a  kh  gh  c  t  ñ  n  r  l  y
    k  g   ṅ   ṭ  p  ṇ  m  ṣ  s  ś

    with all the other letters remaining unchanged.

These codes can serve as sub-units in cryptographic transformations.

**The Kaṭapayādi (Vararuci) Code.** This code has been popular for a long time in India. It consists of mapping of different consonants into the digits 0 through 9. The vowels are free and they may be chosen in a manner so that meaningful words are created.

*Kaṭapayādi Mapping*

1  2  3  4  5  6  7  8  9  0
k  kh g  gh ṅ  c  ch j  jh ñ
ṭ  ṭh ḍ ḍh ṇ  t  th d  dh n
p  ph b  bh m
y  r  l  v  ś  ṣ  s  h

Since the vowels are to be used freely in conjunction with the consonants   1472 = *kava sira*  or  *pava thabha*, etc and likewise 5380 = *mule dana* or *śila hena*.

We now present a *Kaṭapayādi* code for the English alphabet:

*An English Kaṭapayādi Code*

1 2 3 4 5 6 7 8 9 0
b c d f g h j k l m
n p q r s t v w x y
z

Since vowels are free, the pairs of digits

```
45   15   50   11   42   11   13
fg   bs   gm   bn   rp   bn   nd        (one choice)

fog  base game bin  rip  bone nod       (example)
```

## Āryabhaṭa's Substitution Code

Āryabhaṭa's code is a mapping of numbers to words that employs first a division of the Sanskrit alphabet into two groups of consonants and vowels [29].

The first 25 consonants called the *varga* letters (V) of the Sanskrit alphabet are:

*k*=1   *kh*=2   *g*=3    *gh*=4   *ṅ*=5
*c*=6   *ch*=7   *j*=8    *jh*=9   *ñ*=10
*ṭ*=11  *ṭh*=12  *ḍ*=13   *ḍh*=14  *ṇ*=15
*t*=16  *th*=17  *d*=18   *dh*=19  *n*=20
*p*=21  *ph*=22  *b*=23   *bh*=24  *m*=25

The remaining 8 consonants are the *avarga* (A) letters:

*y*=30   *r*=40   *l*=50   *v*=60
*ś*=70   *ṣ*=80   *s*=90   *h*=100

The varga letters (*k* through *m*), V, stand for squares such as 1, 100, 10000, and so on. The avarga letters (*y* through *h*), A, stand for non-squares, such as 10, 1000, and so on. The following vowels are the place holders as shown:

a=1   i=$10^2$   u=$10^4$   ṛ=$10^6$   ḷ=$10^8$
e=$10^{10}$   ai=$10^{12}$   o=$10^{14}$   au=$10^{16}$

This creates a notational system in place values for numbers as large as $10^{17}$:

```
17 16 15 14 13 12 11 10  9  8  7  6  5  4  3  2  1  0
 0  0  0  0  0  0  0  0  0  0  0  0  0  0  0  0  0
au au ai ai  o  o  e  e  ḷ  ḷ  ṛ  ṛ  u  u  i  i  a  a
```

औ औ ऐ ऐ ओ ओ ए ए ॡ ॡ ॠ ॠ उ उ इ इ अ अ

A letter is to be inserted next to its vowel which will be clear by the examples that follow. For more examples, the *Āryabhaṭīya* may be consulted.

**Example 1:**
Message: 3,861

```
A  V  A  V
i  i  a  a
3  8  6  1
y  j  v  k
```

Code =  yijivaka, or kavajiyi or yivakaji
     यिजिवक कवजियि यिवकजि

(permutations of the 4 letters)

**Example 2**: 1,582,237,500 (number of rotations of Earth in a yuga of 432,000 years in Aryabhatiya)
 Vowels and consonants needed:

```
A  V  A  V  V  A  V  A  V
ḷ  ḷ  ṛ  ṛ  u  i  i  a  a
1  5  8  2  2  3  7  5  0  0
ṇ  ṣ  kh 0  b  y  ś  ṅ  0  0
```

Message: 1,583,237,500 = ङिशिबुख़ृष़ृण्ऌ or ण्ऌषृख़ृबुशिङि

Code: *ṅiśibukhṛṣṛṇḷ* or *ṇḷṣṛkhṛbuśiṅi*  or permutation thereof

The table of Rsines in Āryabhaṭa's text is given in verse using the above mapping. This general idea has applications to securing passwords and personal identification numbers as shown by Smith in his patents [37],[38].

**Concluding Remarks**

Āryabhaṭa was one of the greatest mathematician-astronomers of the ancient world who had enormous influence not only in India but also in Western science through the agency of the Arabs. His school was to play an important role in the development of calculus in Kerala.

For computer scientists, the most significant of his contributions is the *kuṭṭaka* algorithm and his substitution code.